\documentclass[
    aps,
    prb,
    twocolumn,
    floatfix,nofootinbib,
    superscriptaddress
]{revtex4-2}

\usepackage{graphicx}
\usepackage{subfigure}
\usepackage[utf8]{inputenc}
\usepackage{times}
\usepackage[breaklinks,colorlinks=true]{hyperref}
\hypersetup{linkcolor=red,citecolor=green,urlcolor=blue}
\usepackage{color}
\usepackage{siunitx}
\usepackage{booktabs}
\usepackage[version=4]{mhchem}

\usepackage{amssymb} 
\usepackage{amsmath}
\usepackage{amsthm}
\usepackage{braket}
\usepackage{MnSymbol}
\usepackage{stmaryrd}
\usepackage{mathtools}

\newcommand{\pgi}{Peter Gr\"unberg Institut and Institute for Advanced Simulation,
Forschungszentrum J\"ulich and JARA, 52425 J\"ulich, Germany}

\newcommand{\aachen}{Department of Physics, RWTH Aachen University, 52056 Aachen, Germany}

\newcommand{\mainz}{Institute of Physics, Johannes Gutenberg University Mainz, 55099 Mainz, Germany}

\newcommand{\abs}{\mathrm{\abs}}

\usepackage{tikz}
\usepackage{tikz-cd}
\usetikzlibrary{calc}
\usetikzlibrary{decorations.markings}
\usepackage{xcolor}
\usepackage{graphicx}

\makeatletter
\renewcommand\@biblabel[1]{#1.}
\makeatother
\begin{document}

\setcounter{secnumdepth}{2} 
%----------------------------------------------------
% Document title
%----------------------------------------------------

\title{Laser-induced charge and spin photocurrents at BiAg$_2$ surface: a first principles benchmark}

\author{T. Adamantopoulos}
    \thanks{t.adamantopoulos@fz-juelich.de}
    \affiliation{\pgi}
    \affiliation{\aachen}

\author{M. Merte}
	%\thanks{m.merte@fz-juelich.de}
    \affiliation{\pgi}
    \affiliation{\aachen}
    \affiliation{\mainz}

\author{D. Go}
    \affiliation{\mainz}
    \affiliation{\pgi}
    
    \author{F. Freimuth}
    \thanks{f.freimuth@fz-juelich.de}
    \affiliation{\mainz}
    \affiliation{\pgi}
    
\author{S. Bl\"ugel}
    \affiliation{\pgi}
    
\author{Y. Mokrousov}
\thanks{y.mokrousov@fz-juelich.de}
    \affiliation{\pgi}
    \affiliation{\mainz}

\date{\today}

\begin{abstract}

%Copied from MMM abstract...The physics of photo-induced effects in interfacial systems is intensively researched these days owing to numerous potential applications. 
%Owing to the complexity of the problem, a comprehensive theoretical picture of photogalvanic effects taking place at realistic metallic surfaces and interfaces is still lacking. 
%In the past, the emergence of charge and spin photocurrents has been shown for the two-band Rashba model, and it was argued that such surface photocurrents stem from interfacial spin-orbit interaction (SOI) which can drive sizeable response when excited by femtosecond laser pulses.
%Additionally, second order spin photocurrents are allowed by symmetry in the non-magnetic Rashba model. 
%These responses stem from the interfacial spin-orbit interaction (SOI) and can be generated by the application of femtosecond laser pulses [1,2]. 
Here, we report first principles calculations and analysis of laser-induced photocurrents at the surface of a prototype Rashba system. By referring to Keldysh non-equilibrium formalism combined with the Wannier interpolation scheme we perform  first-principles electronic structure calculations of a prototype BiAg$_2$ surface alloy, which is a well-known material realization of the Rashba model. 
In addition to non-magnetic ground state situation we also study the case of in-plane magnetized BiAg$_2$. 
%The in-plane magnetization is introduced by the addition of a Zeeman splitting term in the Wannier-interpolated Hamiltonian. 
We calculate the laser-induced charge photocurrents for the ferromagnetic case and the laser-induced spin photocurrents for both the non-magnetic and the ferromagnetic cases. Our results confirm the emergence of very large in-plane photocurrents as predicted by the Rashba model. The resulting photocurrents satisfy all the symmetry restrictions with respect to the light helicity and the magnetization direction. We provide microscopic insights into the symmetry and magnitude of the computed currents based on the {\it ab-initio} multi-band electronic structure of the system, and scrutinize the importance of resonant two-band and three-band transitions for driven currents,  thereby establishing a benchmark picture of photocurrents at Rashba-like surfaces and interfaces. Our work contributes to the study of the role of the interfacial Rashba spin-orbit interaction as a mechanism for the generation of in-plane photocurrents, which are of great interest in the field of ultrafast and terahertz spintronics.  

\end{abstract}

\maketitle

%----------------------------------------------------
% Document body
%----------------------------------------------------

%\tableofcontents %<---- keep for now, remove in the end

%----------------------------------------------------
%\noindent
%{\large{\bf Introduction}}

%\section{Introduction}
%----------------------------------------------------

%----------------------------------------------------

%\maketitle
\date{\today}

%\null\cleardoublepage

%^^^^^^^^^^^^^^^^^^^^^^^^^^^^^^^^^^^^^^^^^^^^^^^^^^^^^^^^^^^^^^^^^^^^^^^^^^^^^^^^^^^^^^^^^^^^^^^^^^^^^^^^^^^^^^^^^^^^^^^^^^^^^^^^^^^^^^^^^^^^^^^^^^^^^^^^^^^^
%^^^^^^^^^^^^^^^^^^^^^^^^^^^^^^^^^^^^^^^^^^^^^^^^^^^^^^^^^^^^^^^^^^^^^^^^^^^^^^^^^^^^^^^^^^^^^^^^^^^^^^^^^^^^^^^^^^^^^^^^^^^^^^^^^^^^^^^^^^^^^^^^^^^^^^^^^^^^
%^^^^^^^^^^^^^^^^^^^^^^^^^^^^^^^^^^^^^^^^^^^^^^^^^^^^^^^^^^^^^^^^^^^^^^^^^^^^^^^^^^^^^^^^^^^^^^^^^^^^^^^^^^^^^^^^^^^^^^^^^^^^^^^^^^^^^^^^^^^^^^^^^^^^^^^^^^^^
%
%					INTRODUCTION
%
%
%
%
%
%
%
%
%
%
%
%
%
%
%\vspace{0.5cm}

%{\it Introduction.} 
\section{Introduction}

%{\color{red}Should we stick to ab-initio, first-principles or keep both?}

The interest in physics of optical generation and properties of laser-induced currents arising at surfaces and interfaces is steadily rising, owing in part to bright prospects that these currents carry for technological applications. Among the latter, the generation of THz radiation with so-called spintronics THz emitters which rely on spinorbitronics effects taking place in magnetic bilayers, have come to occupy a prominent place~\cite{Seifert_2016, PapaioannouBeigang}. As the sources of THz radiation arising from magnetic bilayers exposed to fs laser pulses,
%In the past few years the interest in the field of ultrafast and terahertz (THz) spintronics was piqued by the mechanisms that lie behind the generation of in-plane charge photocurrents which occur after an excitation by fs-laser pulses in magnetic bilayer systems. The motivation behind studying these systems is the occurring THz signals which qualify them as perfect candidates for realising THz emitters~\cite{Seifert_2016, PapaioannouBeigang}. 
two main 
%experimentally identified 
mechanisms of parent in-plane charge currents are considered: generated via the inverse spin Hall effect in response to laser ignited superdiffusive spin-currents~\cite{Kampfrath_2013, Huisman_2017, Battiato_2010, Battiato_2012, Malinowski2008, Melnikov_2011}, and currents generated via the inverse spin-orbit torque in response to the inverse Faraday effect %(IFE) with the mediation of the inverse spin-orbit torque (ISOT)
~\cite{Huisman_2016, Choi_2017, Freimuth_2015}.
However, the diversity of optically mediated planar interfacial currents is not limited to  aforementioned scenarios. For example, recently it was predicted that charge photocurrents can arise from a time-varying exchange splitting following the laser excitation~\cite{Freimuth_2017charge_pumping}. Furthermore, it was pointed out that 
%for an extended two-band ferromagnetic Rashba model~\cite{freimuth_2021} 
laser-induced charge photocurrents can arise in non-centrosymmetric magnetic bilayers as a result of interfacial spin-orbit interaction (SOI)~\cite{freimuth_2021}. This effect was coined as the magnetic photogalvanic effect, to distinguish it from  circular photogalvanic effect which appears in non-centrosymmetric non-magnetic semiconductors~\cite{Ganichev_2003, Ma_2017}. Moreover, it was also predicted that laser-induced spin photocurrents can arise at metallic surfaces, both magnetic as well as non-magnetic~\cite{freimuth_2021,Merte_FGT}. 
%for the non-magnetic, as well as the in-plane ferromagnetic Rashba model.

As one of the fundamental optical phenomena,  photogalvanic effects have been intensively studied in the past and they have gained even more interest in recent years due to their potential applications. However, to date, very little is known about photocurrents emerging at metallic surfaces and interfaces, playing such a central role in THz spintronics. Besides a recent study performed within an effective  Rashba model~\cite{freimuth_2021}, not a single report on first principles calculations of photocurrents in a Rashba-like  system is known to us. This should not come  as too surprising, given a large numerical effort needed for computing non-linear effects from ab-initio, as well as a multitude of discussed sources of photocurrents, whose debated relevance depends on a given system. On the other hand, estimating the relevance of different contributions in a given material from first principles  at times requires completely different numerical approaches, which makes such studies barely feasible. 

Currently, intrinsic contributions which have been identified behind the generation of non-linear optical responses are the shift and injection currents~\cite{KrautBaltz_1981, Sipe_2000, julen_GaAs_pc, Zhang_2018}, the Berry curvature dipole (BCD) term~\cite{Sodemann_2015, Sodemann_2019} as well as the semiclassical Jerk~\cite{Sodemann_2019, Fregoso_2018, Fregoso_2019} and ballistic terms~\cite{Sturman_2020}. The shift and injection terms are important in gapped materials and on this front first-principles calculations were performed for example in ferroelectric materials~\cite{KrautBaltz_1981, Mu_2021}, semiconductors~\cite{Sipe_2000, julen_GaAs_pc}, quantum wells~\cite{Sherman_2005}, and graphene~\cite{Yin_2019}. In metals, the presence of the Fermi surface gives rise to the BCD and semiclassical terms, also referred to as metallic terms. Especially for the BCD, it was proposed that the non-linear Hall effect can be observed in materials with large Berry curvature stemming from accidental or avoided band-crossings like topological crystalline insulators, two-dimensional transition metal dichalcogenides, and three-dimensional Weyl semimetals~\cite{Sodemann_2015}. Recently, we developed an \textit{ab-initio} scheme to calculate second order optical response properties  within the Keldysh formalism~\cite{freimuth_2016, freimuth_2021}, successfully applying it to study charge and photocurrents in single-layer ferromagnetic Fe$_3$GeTe$_2$~\cite{Merte_FGT}. 
%This formula was already tested in the calculation of charge and spin photocurrents in the two-dimensional ferromagnetic single-layer Fe$_3$GeTe$_2$~\cite{Merte_FGT}
The developed approach is optimal in providing reference values in realistic materials, as it can treat both insulating and metallic systems on equal footing, and it naturally lends itself to including various disorder effects, without the need to distinguish among intrinsic and various extrinsic contributions to the photocurrents explicitly.

In this work, we choose a prototype metallic Rashba system $-$ BiAg$_2$ surface alloy~\cite{Carbone_2016} $-$ and apply the developed Keldysh methodology to study the properties of charge and spin photocurrents at it from first principles. With this, we aim to provide benchmark values and   develop a material-specific theory of photocurrents at metallic magnetic and non-magnetic Rashba surfaces, which can be used in the future as a reference point by other ab-initio studies. Specifically,  
%Motivated by the predictions of the ferromagnetic Rashba model~\cite{freimuth_2021}, in this work we employ the Keldysh non-equilibrium formula and calculate the \textit{ab-initio} laser-induced in-plane charge and spin photocurrents which arise in a BiAg$_2$ surface. 
we study the properties of photocurrents in the non-magnetic ground state and in an in-plane magnetized ferromagnetic case, analyzing our results in terms of disorder strength, band filling and frequency of the light. Apart from confirming the predictions of the Rashba model we also highlight the importance of the Rashba splitting and exchange interaction in generating the responses stemming from resonant two-band transitions in the ferromagnetic case. 
As such, we also build an intuition in  required material parameters for strong optical response, necessary for experimental materials engineering and promising applications in the field of ultrafast and THz spintronics. Our work is structured as follows. In Section~\ref{Method} we briefly describe the computational methodology, and in Section~\ref{Computational Details} we provide structural details and details of first principles calculations. In Section~\ref{Ferromagnetic charge photocurrents} we address the  properties of charge photocurrents in the ferromagnetic case, while in Section~\ref{Non-magnetic spin photocurrents} and~\ref{Ferromagnetic charge photocurrents} we study the physics of spin photocurrents at the non-magnetic and magnetic surface, respectively. In Section~\ref{two-band_vs_three-band} we discuss the relevance of different electronic transitions behind the computed photocurrents. Our manuscript ends with a summary in Section\ref{Summary}.

\section{Method}\label{Method}

In this work we calculate the charge and spin photocurrents which arise at second order in the perturbing electric field of a continuous laser pulse of frequency $\omega$ by using the expressions which were previously derived in~\cite{freimuth_2021} within the framework of the Keldysh formalism. The expression for the second order charge photocurrents flowing in direction $i$ is~\cite{freimuth_2021}:
%In order to compute the photocurrents in the system arising as a response to a continuous laser pulse of frequency $\omega$, we employ an expression for the second order photocurrent density which was previously derived by us using Keldysh formalism
%(to which we refer as the current for shortness of notation)~
%~\cite{freimuth_2021}:
\begin{equation}
    J_{i}=\frac{a_{0}^{2} e I}{\hbar c}\left(\frac{\mathcal{E}_{\mathrm{H}}}{ \hbar \omega}\right)^{2} \operatorname{Im} \sum_{j k} \epsilon_{j} \epsilon_{k}^{*} \varphi_{i j k} ,
    \label{eq:photocurrent}
\end{equation}
%\textbf{Suggestion FF: Alternative Equation without intensity and field amplitudes instead of polarizations:}
%\begin{equation}
%    J_{i}=\frac{a_{0}^{2} e \epsilon_{0}}{2\hbar }\left(\frac{\varepsilon_{\mathrm{H}}}{ \hbar \omega}\right)^{2} \operatorname{Im} \sum_{j k} E_{j} E_{k}^{*} \varphi_{i j k},
%    \label{eq:photocurrent}
%\end{equation}
where the quantity $\varphi_{ijk}$
%, to which we refer below as the second order conductivity tensor,
has the form:
\begin{equation}
    \begin{aligned}
\varphi_{i j k}=& \frac{2}{a_{0} \mathcal{E}_{\mathrm{H}}} \int \frac{\mathrm{d}^{2} k}{(2 \pi)^{2}} \int \mathrm{d} \mathcal{E} \operatorname{Tr}[\\
& f(\mathcal{E}) v_{i} G_{\boldsymbol{k}}^{\mathrm{R}}(\mathcal{E}) v_{j} G_{\boldsymbol{k}}^{\mathrm{R}}(\mathcal{E}-\hbar \omega) v_{k} G_{\boldsymbol{k}}^{\mathrm{R}}(\mathcal{E}) \\
-& f(\mathcal{E}) v_{i} G_{\boldsymbol{k}}^{\mathrm{R}}(\mathcal{E}) v_{j} G_{\boldsymbol{k}}^{\mathrm{R}}(\mathcal{E}-\hbar \omega) v_{k} G_{\boldsymbol{k}}^{\mathrm{A}}(\mathcal{E}) \\
+& f(\mathcal{E}) v_{i} G_{\boldsymbol{k}}^{\mathrm{R}}(\mathcal{E}) v_{k} G_{\boldsymbol{k}}^{\mathrm{R}}(\mathcal{E}+\hbar \omega) v_{j} G_{\boldsymbol{k}}^{\mathrm{R}}(\mathcal{E}) \\
-& f(\mathcal{E}) v_{i} G_{\boldsymbol{k}}^{\mathrm{R}}(\mathcal{E}) v_{k} G_{\boldsymbol{k}}^{\mathrm{R}}(\mathcal{E}+\hbar \omega) v_{j} G_{\boldsymbol{k}}^{\mathrm{A}}(\mathcal{E}) \\
+& f(\mathcal{E}-\hbar \omega) v_{i} G_{\boldsymbol{k}}^{\mathrm{R}}(\mathcal{E}) v_{j} G_{\boldsymbol{k}}^{\mathrm{R}}(\mathcal{E}-\hbar \omega) v_{k} G_{\boldsymbol{k}}^{\mathrm{A}}(\mathcal{E}) \\
+&\left.f(\mathcal{E}+\hbar \omega) v_{i} G_{\boldsymbol{k}}^{\mathrm{R}}(\mathcal{E}) v_{k} G_{\boldsymbol{k}}^{\mathrm{R}}(\mathcal{E}+\hbar \omega) v_{j} G_{\boldsymbol{k}}^{\mathrm{A}}(\mathcal{E})\right] .
\end{aligned}
\label{eq:conductivity}
\end{equation}
In the above expressions $a_0$ is the Bohr's radius, $e$ is the elementary charge, $I$ is the intensity of the pulse, $\hbar$ is the reduced Planck constant, $c$ is the light velocity, $\mathcal{E}_H=e^2/(4\pi \epsilon_0 a_0)$ is the Hartree energy, and $\epsilon_j$ is the $j$'th component of the polarization vector of the pulse. With $v_{i}$ we label the $i$'th component of the velocity operator, $f(\mathcal{E})$ is the Fermi-Dirac distribution function and $G_{\boldsymbol{k}}^{\mathrm{R(A)}}$ stand for the retarded (advanced) Green function of the system. The expression for the spin photocurrent $J^{s}_{i}$ flowing in direction $i$ with spin polarization along axis $s$, is obtained by replacing the first of the velocity operators $v_i$ in Eq.~\eqref{eq:conductivity}, with the operator of the spin velocity $\{v_i,\sigma_s\}$, and the prefactor $a_{0}^{2} e I / \hbar c$ Eq.~\eqref{eq:photocurrent} with the prefactor $-a_{0}^{2} I / 4 c$.

% where $a_0$ is the Bohr's radius, $e$ is the elementary charge, $\hbar$ is the reduced Planck constant, $c$ is the speed of light, $\varepsilon_H=e^2/(4\pi \epsilon_0 a_0)$ is he Hartree energy, $\epsilon_i$ is the $i$'th component of the polarization vector of the pulse, and $I$ is the pulse intensity. 
 %The field intensity is given by $I=\epsilon_0 c E^2_0/2$, where $\epsilon_0$ is the vacuum permittivity.

The energy dependent Green function of the system is given by~\cite{freimuth_2021}:
\begin{equation}
    G_{\boldsymbol{k}}^{\mathrm{R}}(\mathcal{E})=\hbar \sum_{n} \frac{|\boldsymbol{k} n\rangle\langle \boldsymbol{k} n|}{\mathcal{E}-\mathcal{E}_{\boldsymbol{k} n}+i \Gamma} \ \ \text{and} \ \  G_{\boldsymbol{k}}^{\mathrm{A}}(\mathcal{E})= [ G_{\boldsymbol{k}}^{\mathrm{R}}(\mathcal{E}) ]^\dagger,
\label{eq:green}
\end{equation}
where $\mathcal{E}_{\boldsymbol{k} n}$ is the energy of the state $\Ket{\boldsymbol{k}n}$ with band index $n$ and a Bloch vector $\boldsymbol{k}$. In order to describe the effect of disorder of the electronic states, the constant lifetime broadening $\Gamma$ is introduced. The parameter $\Gamma$ allows us to perform the energy integration in Eq.~\eqref{eq:conductivity} analytically in the way described in~\cite{freimuth_2016}.

%In this work, we model the effect of disorder by adapting a model of constant lifetime broadening of the states $\Gamma$ which results in the following expressions: $G_{\boldsymbol{k}}^{\mathrm{R}}(\mathcal{E})=\hbar \sum_{n} \frac{|\boldsymbol{k} n\rangle\langle \boldsymbol{k} n|}{\mathcal{E}-\mathcal{E}_{\boldsymbol{k} n}+i \Gamma}$ and  $ G_{\boldsymbol{k}}^{\mathrm{A}}(\mathcal{E})= [ G_{\boldsymbol{k}}^{\mathrm{R}}(\mathcal{E}) ]^\dagger$~\cite{freimuth_2021}, where the energy of the state  $\Ket{\boldsymbol{k}n}$ in  band $n$ with a Bloch vector $\boldsymbol{k}$ is $\mathcal{E}_{\boldsymbol{k} n}$.
%To compute the photocurrent the integrals in Eq.~\eqref{eq:conductivity} have to be evaluated. At zero temperature the Fermi distribution becomes a step function which allows one to perform the energy integration analytically
%and corresponding analytic expressions can be found in Appendix B of  
%Ref.~\cite{freimuth_2016}. The numerical evaluation is preformed within the basis of maximally localized Wannier functions, and the  Brillouin zone integration  is performed numerically by employing the efficient technique of Wannier interpolation~\cite{code_w90,code_fleurWann,code_w90_ahc}. A more detailed description of the methodology will be published elsewhere. Throughout this work we assume an intensity of the light of 10\,GW/cm$^2$, which corresponds to typical values of the fluence of the order of 0.5\,mJ/cm$^2$ for a 50\,fs laser pulse~\cite{huisman_2016}.

A circularly polarised pulse propagating in the $z$ direction is described by $\boldsymbol{\epsilon}=(1,\lambda i,0)/\sqrt{2}$, where $\lambda=\pm 1$ is the helicity. Linearly polarised light along $x$ or $y$ axis is described by $\boldsymbol{\epsilon}=(1,0,0)$ and $\boldsymbol{\epsilon}=(0,1,0)$, respectively. The assumed intensity of the pulse is 10\,GW/cm$^2$, which corresponds to typical values of the fluence of the order of 0.5\,mJ/cm$^2$ for a 50\,fs laser pulse~\cite{Huisman_2016}.

%{\it Computational Details.}

\section{Computational Details}\label{Computational Details}

\begin{figure}[t!]
\begin{center}
\rotatebox{0}{\includegraphics [width=0.95\columnwidth]{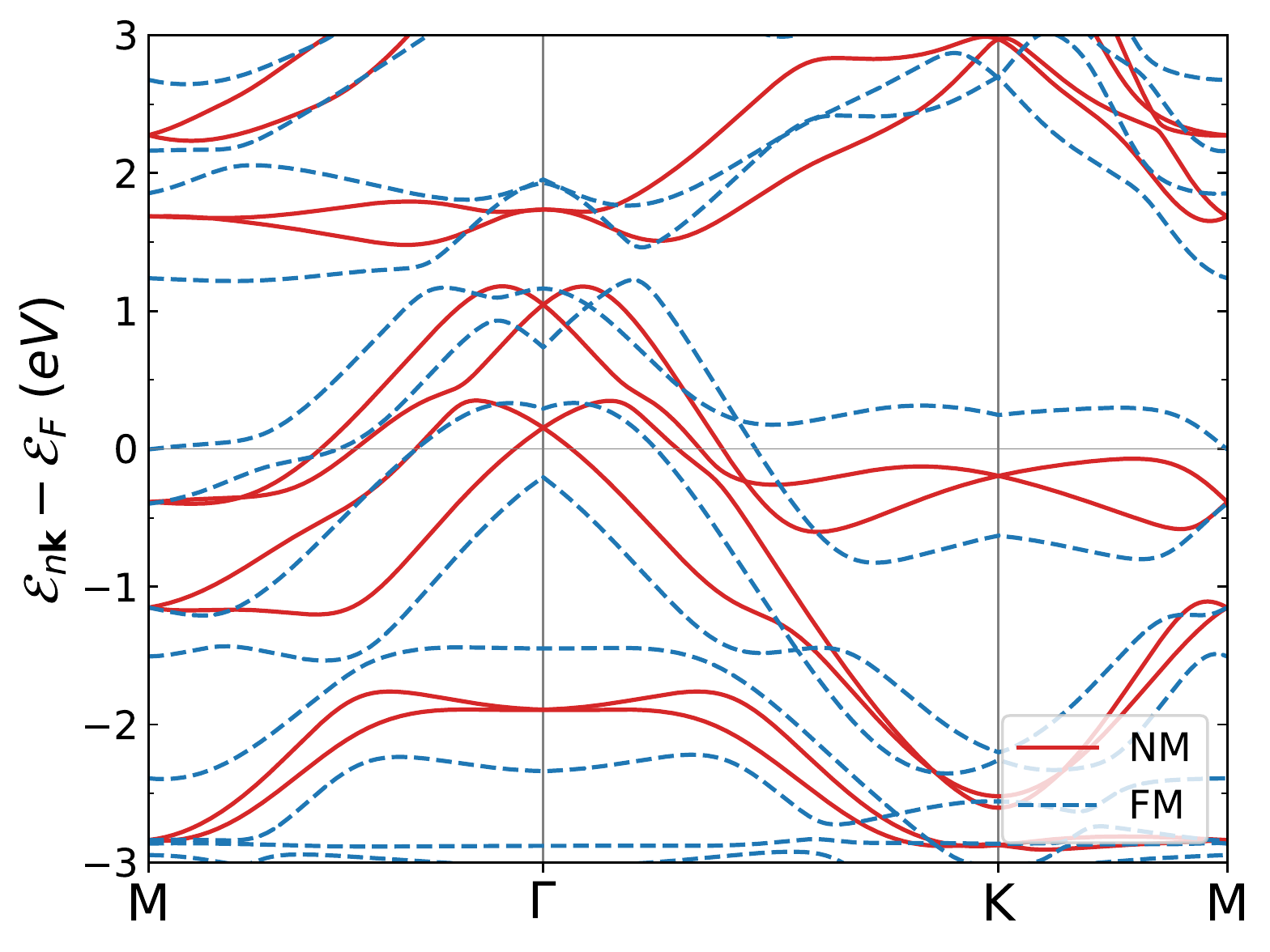}}
\end{center}
\caption{The interpolated band-structure of  BiAg$_2$  around the Fermi energy  for the non-magnetic (NM, solid red line) and the ferromagnetic (FM, dashed blue line) case. For the ferromagnetic case an exchange field with a strength of 0.5\,eV is applied in-plane along the $y$-axis.}%
\label{BiAg2_structure}
\end{figure}

The electronic structure of the BiAg$_2$ surface was calculated from first-principles by using the film mode of the full-potential linearized augmented plane wave \texttt{FLEUR} code~\cite{fleur}. The spin-orbit coupling (SOC) effect was treated within the second-variation scheme~\cite{SOC_2nd_var}. For the exchange and correlation effects the non-relativistic PBE~\cite{pbe} functional was used. The in-plane lattice constant was chosen at $a=9.466$\,a.u, while the surface relaxation of the Bi atom was set to $d=1.61$\,a.u. $-$
the values determined from our previous first principles calculations~\cite{dgo_srep}. The muffin-tin radii of Bi and Ag atoms were set to 2.80\,a.u$^{-1}$ and 2.59\,a.u$^{-1}$, respectively. The   plane-wave cutoff was set to $K_{\text{max}}=4.0$\,a.u$^{-1}$ and we used a set of 128 $k$-points in the  Brillouin zone for self-consistent calculations. The crystal and electronic structure of the system can be seen in Fig. 1 of Ref.~\cite{dgo_srep}. From the first-principles band structure the Rashba-like effect of including SOC on  the $k$-dependent band splitting of the states is evident, see also Fig.~\ref{BiAg2_structure}.

Next, by using the Wannier90 code~\cite{Pizzi2020} we obtained maximally-localized Wannier functions (MLWFs) which reproduce the first-principles electronic structure up to a frozen window of 2.78\,eV above the Fermi energy. The chosen initial projections were $sp_2$ and $p_z$ orbitals for Bi, while for Ag they were $s$, $p$ and $d$ orbitals. In this way we constructed 44 MLWFs out of 120 Bloch functions (BFs) on a 16$\times$16 $k$-mesh.

\begin{figure*}[ht!]
\begin{center}
\rotatebox{0}{\includegraphics [width=0.94\linewidth]{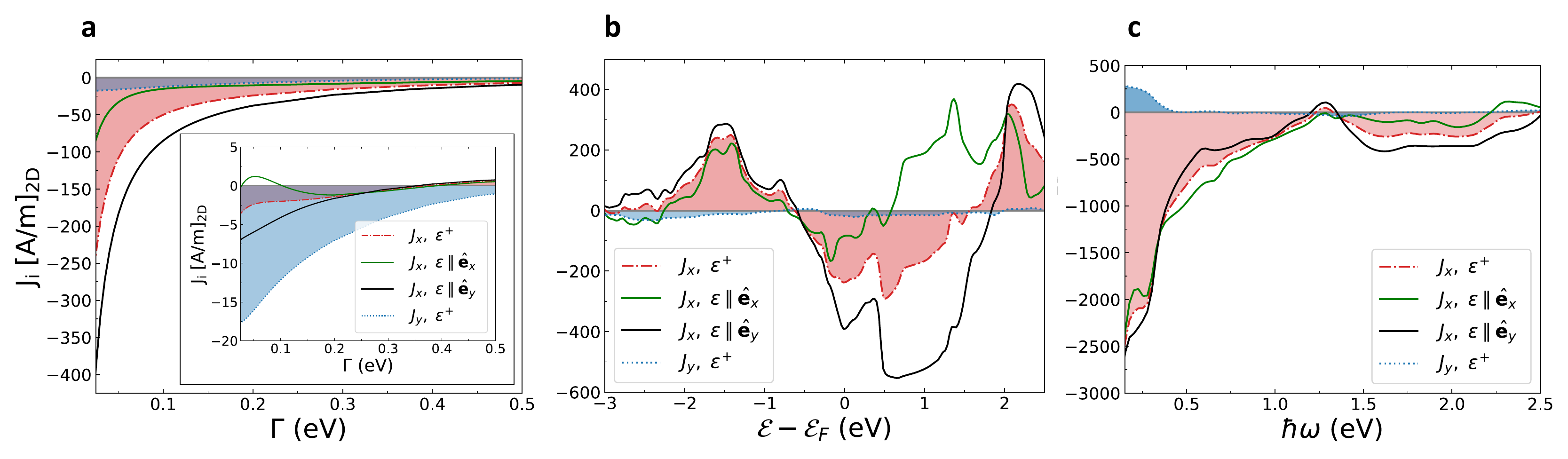}}
\end{center}
\caption{Charge photocurrents in relation to (\textbf{a}) the lifetime broadening $\Gamma$, (\textbf{b}) the Fermi energy level $\mathcal{E}_F$, and (\textbf{c}) the frequency of  light  $\hbar\omega$ for the ferromagnetic BiAg$_2$ surface. The shown symmetry-allowed components are $J_x$ for circularly polarised light (shadowed dashed red line) and linearly polarised light along  $x$- (green line) or $y$-axis (black line), as well as $J_y$ for circularly polarised light (shadowed dotted blue line). The inset in (\textbf{a}) depicts only the three-band contributions to the ferromagnetic charge photocurrents,  discussed in Sec.~\ref{two-band_vs_three-band}. In (\textbf{a}, \textbf{b}) $\hbar \omega=1.55$\,eV and in (\textbf{b}, \textbf{c}) $\Gamma=25$\,meV. In (\textbf{a}, \textbf{c}) the calculations are performed for true Fermi energy. For all calculations the considered light intensity is $I=10$\,GW/cm$^2$.}
\label{FM_charge_photoc}
\end{figure*}
%\vspace{0.5cm}

In a post-processing step the interpolated Hamiltonian was constructed  out of the previously obtained MLWFs. In its ground state the system is non-magnetic, and starting from non-magnetic MLWFs aided by the matrix elements of Pauli matrices~\cite{Zhang_2012}, an exchange field of the type
\begin{equation}
    H_{ex}=\frac{\Delta V}{2} \boldsymbol{\sigma} \cdot \hat{\mathbf{n}}(\mathbf{r})
\end{equation}
was applied on the interpolated Hamiltonian in order to make the system ferromagnetic and break the time-reversal symmetry. Following the symmetry  analysis, performed within the the  Rashba model and presented in Ref.~\cite{freimuth_2021}, which predicts vanishing photo-signal for the case of out-of-plane magnetization,   we choose the direction of the exchange field to be in-plane along the $y$-axis. The calculated band structure of the ferromagnetic case presented in comparison to the non-magnetic case is shown  in Fig.~\ref{BiAg2_structure}. Additional asymmetric splittings of the bands, as well as the shift of the metallic Weyl point positioned at $+1$\,eV,  characteristic for the Rashba model, away from the $\Gamma$-point arising in response to the applied exchange field  are clearly visible~\cite{Carbone_2016}.

In the following, we compute the photocurrents in the system and vary the lifetime broadening $\Gamma$ in a region of $[0.025, 0.5]$\,eV, the light energy $\hbar\omega$ between $[0.15, 2.5]$\,eV and we cover an energy region of $[-3, 2.5]$\,eV around the Fermi energy level $\mathcal{E}_F$. A 512$\times$512 $\mathbf{k}$-mesh has proved to be sufficient to obtain converged results. The strength of the exchange field was set at $\Delta V=0.5$\,eV.

%   SUBSEC: WANNIER INTERPOLATION FROM FLEUR

%The Wannier interpolated velocity operator is given by \cite{code_w90_ahc}
%\begin{equation}
 %   v_{n m, \alpha}^{(\mathrm{H})}=\frac{1}{\hbar} \bar{H}_{n m, \alpha}^{(\mathrm{H})}-\frac{i}{\hbar}\left(\mathcal{E}_{m}^{(\mathrm{H})}-\mathcal{E}_{n}^{(\mathrm{H})}\right) \bar{A}_{n m, \alpha}^{(\mathrm{H})},
%\end{equation}
%where  $ \bar{H}_{n m, \alpha}$ is the $\mathbf{k}$-derivative of the Hamiltonian, $\mathcal{E}_{n}$ the eigenvalue of the $n$ th Wannier function and $ \bar{A}_{n m, \alpha}^{(\mathrm{H})}$ is the Berry connection. 

%\subsection{Symmetry analysis}

%{\it Symmetry analysis.}
%\section{Symmetry analysis}

%^^^^^^^^^^^^^^^^^^^^^^^^^^^^^^^^^^^^^^^^^^^^^^^^^^^^^^^^^^^^^^^^^^^^^^^^^^^^^^^^^^^^^^^^^^^^^^^^^^^^^^^^^^^^^^^^^^^^^^^^^^^^^^^^^^^^^^^^^^^^^^^^^^^^^^^^^^^^
%^^^^^^^^^^^^^^^^^^^^^^^^^^^^^^^^^^^^^^^^^^^^^^^^^^^^^^^^^^^^^^^^^^^^^^^^^^^^^^^^^^^^^^^^^^^^^^^^^^^^^^^^^^^^^^^^^^^^^^^^^^^^^^^^^^^^^^^^^^^^^^^^^^^^^^^^^^^^
%^^^^^^^^^^^^^^^^^^^^^^^^^^^^^^^^^^^^^^^^^^^^^^^^^^^^^^^^^^^^^^^^^^^^^^^^^^^^^^^^^^^^^^^^^^^^^^^^^^^^^^^^^^^^^^^^^^^^^^^^^^^^^^^^^^^^^^^^^^^^^^^^^^^^^^^^^^^^
%
%				RESULT SECTION
%
%
%
%
%
%
%
%~~~~~~~~~~~~~~~~
%~~~~~~~~~~~~~~~~
%~~~~~~~~~~~~~~~~
%~~~~~~~~~~~~~~~~
%~~~~~~~~~~~~~~~~
%~~~~~~~~~~~~~~~~
%~~~~~~~~~~~~~~~~

%\cleardoublepage

%{\it Results.}

\section{Results}

The results of our calculations and analysis regarding the charge and spin photocurrents are presented in the following section. In general there is an agreement between the symmetry of the photocurrents that we observe from calculations and the symmetry restrictions derived from the effective  Rashba model in~Ref.~\cite{freimuth_2021}. For simplicity we present the results only for the allowed by symmetry  non-zero components. Furthermore, for circularly polarised light we present only the results for the case of positive helicity $\lambda=+1$, while the case with  $\lambda=-1$  can be reconstructed from the symmetry properties presented in~\cite{freimuth_2021}.

\subsection{Ferromagnetic charge photocurrents}\label{Ferromagnetic charge photocurrents}

We start by presenting in Fig.~\ref{FM_charge_photoc}(\textbf{a}) the results of our calculations for the laser induced charge photocurrents at the ferromagnetic BiAg$_2$ surface with in-plane magnetization along the $y$-axis, in relation to the lifetime broadening of the electronic states $\Gamma$. The calculations are performed at the true Fermi energy level $\mathcal{E}_F$ for a light frequency of $\hbar\omega=1.55$\,eV. We observe that the largest response arises for $J_x$ from linearly polarised light along the $y$-axis (black line) with a magnitude of nearly 400\,A/m at $\Gamma=25$\,meV. The second largest response is for $J_x$ driven by circularly polarised light (shadowed dashed red line) with a magnitude of approximately 225\,A/m at $\Gamma=25$\,meV.
The remaining two types of photocurrent $-$ $J_x$ from light linearly polarised along the $x$-axis (green line) and 
$J_y$ from circularly polarised light (shadowed dotted blue line) $-$
are significantly smaller in magnitude, reaching a range of values from $-$25 to $-$75\,A/m for small values of $\Gamma$.
%The third larger response with a magnitude of 75\,A/m at $\Gamma$=25\,meV is for $J_x$ with light linearly polarised along the $x$-axis (green line). The smallest response arises for $J_y$ with circularly polarised light (shadowed dotted blue line) and a magnitude of nearly 25\,A/m at $\Gamma$=25\,meV. 

In comparison to the results obtained from the effective ferromagnetic Rashba model as presented in Fig.2 of Ref.~\cite{freimuth_2021}, the values presented here are two orders of magnitude larger for the same laser pulse parameters. A possible source of this difference can lie in  different values of the Rashba parameter used. In~\cite{freimuth_2021} a value of 0.1\,eV\r{A} was used to model a Co/Pt magnetic bilayer, while for BiAg$_2$(111) surface alloys the experimentally measured value is about 3.05\,eV\r{A}~\cite{BiAg2(111)_surf_alloys}. This highlights the role of the interfacial Rashba SOI strength in generating surface  photocurrents.
Moreover, as we see in Fig.~\ref{FM_charge_photoc}(\textbf{a}), the charge photocurrents decrease rapidly as the lifetime broadening $\Gamma$ increases. For the values of $\Gamma$ below 0.2\,eV photocurrents exhibit a highly non-linear behavior, while for $\Gamma$ above 0.2\,eV the latter behavior can be very well described by a  $\sim\frac{1}{\Gamma}$ functional form. This kind of dependence of the photocurrents on the lifetime broadening was also recently observed for single-layer Fe$_3$GeTe$_2$~\cite{Merte_FGT} by employing the same computational technique.
%, where the lack of a clear description was attributed to the non-linear nature of the effect.

The magnitude and direction of the photocurrents also depend  very strongly on the position of the Fermi energy. In Fig.~\ref{FM_charge_photoc}(\textbf{b}) we present the results of  calculated charge photocurrents as a function of the position of the Fermi energy in the electronic structure  $\mathcal{E}_F$ while keeping $\Gamma=25$\,meV and $\hbar\omega=1.55$\,eV. Concerning the magnitude, generally similar behavior to that in Fig.~\ref{FM_charge_photoc}(\textbf{a}) is observed, where the response of $J_x$ to light with linear polarization along the $y$-axis is the largest. An exception to this appears in the region of $[+1,+2]$\,eV where a change of sign occurs and the response of $J_x$ for light with linear polarization along the $x$-axis becomes dominant. Note that $J_y$ signal is strongly suppressed in the entire range of band filling. Qualitatively, the shape of $J_x$ response is similar for each situation with some pronounced variations in the signal appearing which cause strong and sharp variation in the signal. Such variations can be noticed, for example, around $-1.5$\,eV, around 0\,eV and around $+1.25$\,eV. By comparing with the bandstructure presented in Fig.~\ref{BiAg2_structure}, we can identify there regions as the regions where bands become flatter or the number of participating in photocurrent bands increases. Therefore when the Fermi energy falls into these regions, the number of ``activated" electronic transitions increases leading to increased magnitude of the photocurrents.
%is locally increased and setting $\mathcal{E}_F$ in the neighborhood of these regions results in larger photocurrents.

\begin{figure*}[t!]
\begin{center}
\rotatebox{0}{\includegraphics [width=0.94\linewidth]{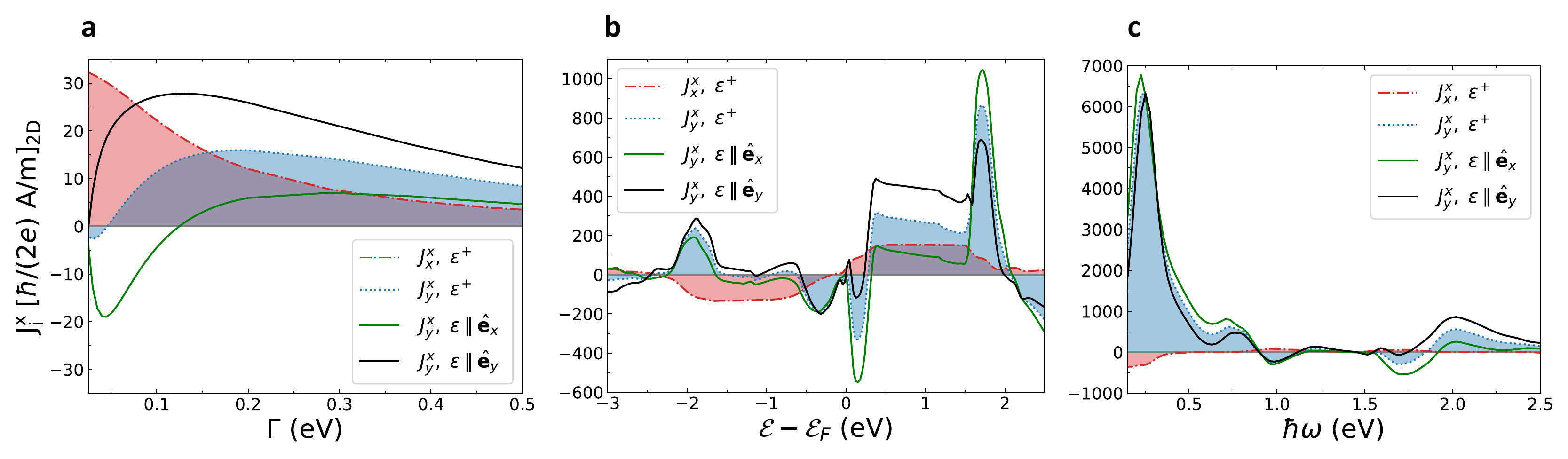}}
\end{center}
\caption{Spin photocurrents in relation to (\textbf{a}) the lifetime broadening $\Gamma$, (\textbf{b}) the Fermi energy level $\mathcal{E}_F$, and (\textbf{c}) the light frequency $\hbar\omega$ for the non-magnetic BiAg$_2$ surface. The symmetry allowed components which are shown are the $J^x_x$ for circularly polarised light (shadowed dashed red line), as well as $J^x_y$ for circularly polarised light (shadowed dotted blue line) and linearly polarised light along the $x$ (green line) or $y$ (black line) axis. In (\textbf{a}, \textbf{b}) $\hbar \omega=1.55$\,eV and in (\textbf{b}, \textbf{c}) $\Gamma=25$\,meV. In (\textbf{a}, \textbf{c}) the calculations are performed for the true Fermi energy. For all calculations the considered light intensity is $I=10$\,GW/cm$^2$.}
\label{NM_spin_photoc}
\end{figure*}
%\vspace{0.5cm}

The dependence of charge photocurrents on the laser frequency $\hbar\omega$ is shown in Fig.~\ref{FM_charge_photoc}(\textbf{c}). In this plot the Fermi energy is set to the true Fermi level of BiAg$_2$, and a broadening of $\Gamma=25$\,meV is used. Concerning the magnitude of $J_x$, the resulting responses exhibit a non-linear behavior in the region below 1.25\,eV and a more stable behavior at higher frequencies. In general the response for linearly polarised light along the $y$-axis (black line) is the largest, except for the region of $[0.5,1.0]$\,eV where the response to linearly polarised light along the $x$-axis (green line) becomes dominant. An interesting change of sign appears at an energy of around 1.25\,eV for the $J_x$ responses. The $J_y$ response is, on the other hand, significant only in a relatively small frequency range around zero.

\subsection{Non-magnetic spin photocurrents}\label{Non-magnetic spin photocurrents}

While the charge photocurrents are vanishing for non-magnetic BiAg$_2$ due to symmetry, this is not the case for spin currents. We move on to present the results for the laser induced spin photocurrents for the non-magnetic BiAg$_2$ surface in Fig~\ref{NM_spin_photoc}. We present only the $J^{x}_{i}$ responses with spin polarization along the $x$-axis. The symmetry allowed $J^{y}_{i}$ responses arise  as equal in magnitude and either the same or opposite in sign 
%as even or odd functions of the 
to $J^{x}_{i}$ ones, and thus we omit them in the figure for simplicity. For circularly polarised light, $J^{x}_{x}$ and $J^{y}_{y}$ are the same, while $J^{x}_{y}$ is opposite to $J^{y}_{x}$. For linearly polarised light, $J^{x}_{y}$ polarised along $x$ is opposite to $J^{y}_{x}$ polarised along $y$, while $J^{x}_{y}$ polarised along $y$ is opposite to $J^{y}_{x}$ polarised along $x$. The above relations are summarized in Table~\ref{NM_spin_photoc_table}.

\begin{table}[b!]
\centering
\begin{tabular}{||c | c ||}
 \hline
 $J^{x}_{i}$ & $J^{y}_{i}$ \\ [0.5ex] 
 \hline\hline
 %\hline
 $J^{x}_{x}$, $\epsilon^+$ & \ \ \ $J^{y}_{y}$, $\epsilon^+$ \\ [0.5ex]
 \hline
 $J^{x}_{y}$, $\epsilon^+$ & \ $-J^{y}_{x}$, $\epsilon^+$ \\ [0.5ex]
 \hline
 $J^{x}_{y}$, $\epsilon \parallel \hat{\mathbf{e}}_x$ & \ $-J^{y}_{x}$, $\epsilon \parallel \hat{\mathbf{e}}_y$ \\ [0.5ex]
 \hline
 $J^{x}_{y}$, $\epsilon \parallel \hat{\mathbf{e}}_y$ & \ $-J^{y}_{x}$, $\epsilon \parallel \hat{\mathbf{e}}_x$ \\ [0.5ex]
 \hline
\end{tabular}
\caption{Relation between the symmetry allowed $J^{x}_{i}$ and $J^{y}_{i}$ current responses of the spin photocurrents at  non-magnetic  BiAg$_2$ surface. With $\epsilon^+$ we denote circularly polarised light with $\lambda=+1$ helicity, while with $\epsilon \parallel \hat{\mathbf{e}}_x(\hat{\mathbf{e}}_y)$ we denote linearly polarised light along $x(y)$.}
\label{NM_spin_photoc_table}
\end{table}

In Fig.~\ref{NM_spin_photoc}(\textbf{a}) the results with respect to the lifetime broadening $\Gamma$ are presented. The calculations were performed at the true Fermi energy $\mathcal{E}_F$ for the laser pulse frequency of $\hbar\omega=1.55$\,eV. For the $J^x_x$ response for circularly polarised light (shadowed dashed red line) we observe the same behavior as in the case of charge photocurrents,~i.e.~non-linear below 0.2\,eV and more stable, $1/\Gamma$-like, above 0.2\,eV. At $\Gamma=25$\,meV the calculated $J^x_x$ has a magnitude of around 30\,$\hbar$/(2e)\,A/m. The $J^x_y$ responses exhibit a different behavior. At small values of $\Gamma$ they converge to zero after reaching a peak value, while for larger values of $\Gamma$ they exhibit a steady $1/\Gamma$ behavior. The current response to circularly polarised light (shadowed dotted blue line) peaks at a magnitude of 15\,$\hbar$/(2e)\,A/m for $\Gamma\approx0.2$\,eV. In the case of linearly polarised light, for the polarization along $x$ (green line) the peaked value is $-20$\,$\hbar$/(2e)\,A/m at $\Gamma\approx 50$\,meV, while for  polarization along $y$ (black line) the peaked value is nearly $30\,\hbar$/(2e)\,A/m at $\Gamma\approx 0.12$\,eV.

\begin{figure*}[ht!]
\begin{center}
\rotatebox{0}{\includegraphics [width=0.9\linewidth]{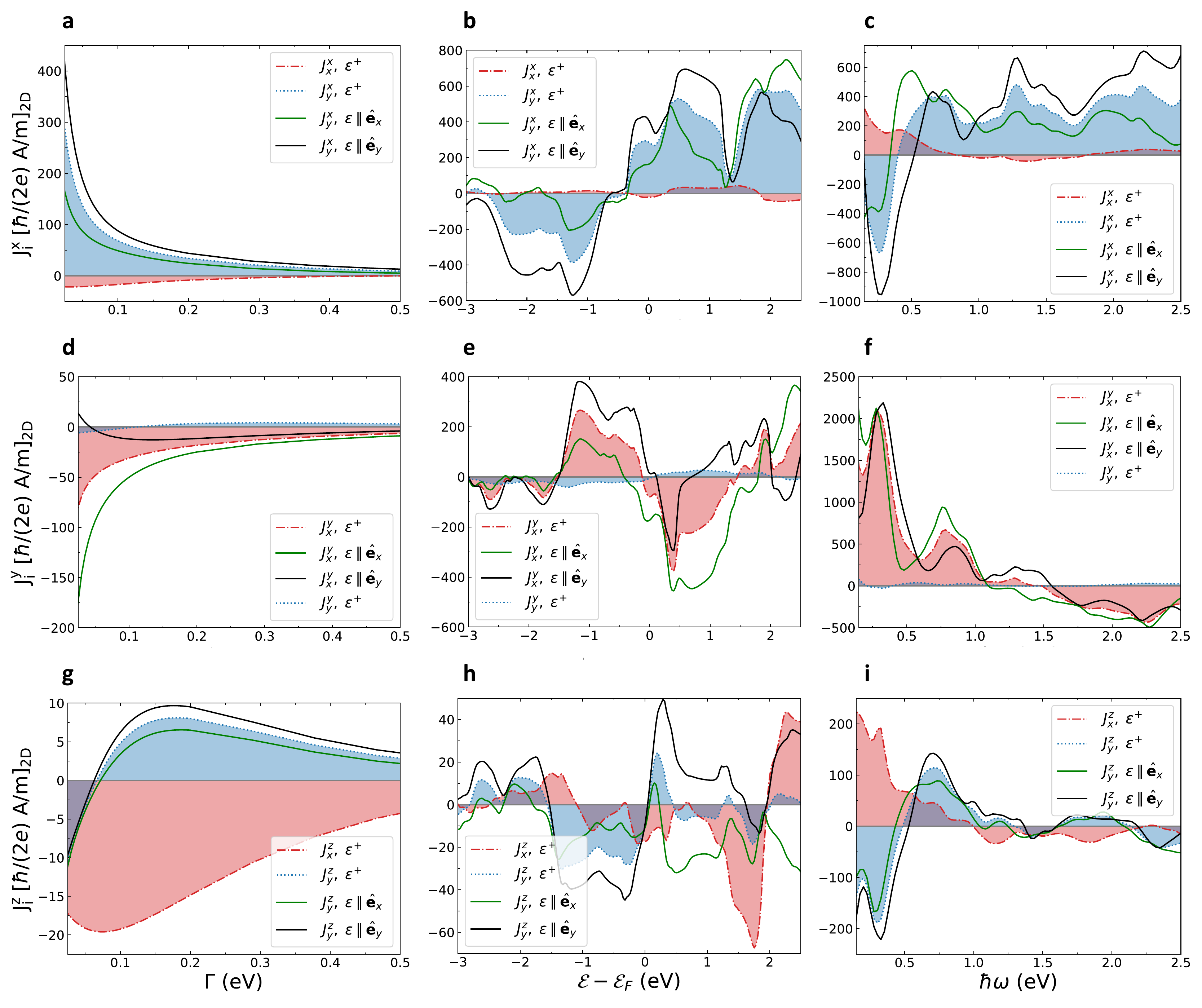}}
\end{center}
\caption{Spin photocurrents in relation to (\textbf{a}, \textbf{d}, \textbf{g}) the lifetime broadening $\Gamma$, (\textbf{b}, \textbf{e}, \textbf{h}) the Fermi energy level $\mathcal{E}_F$, and (\textbf{c}, \textbf{f}, \textbf{i}) the light frequency $\hbar\omega$ for the ferromagnetic BiAg$_2$ surface. In (\textbf{a}-\textbf{c}) $J^x_x$ and $J^x_y$ with circularly polarised light (shadowed dashed red and shadowed dotted blue lines, respectively) together with $J^x_y$ from linearly polarised along the $x$ (green line) or $y$ (black line) axis light are shown. In (\textbf{d}-\textbf{f}) $J^y_x$ and $J^y_y$ from circularly polarised light (shadowed dashed red and shadowed dotted blue lines, respectively) together with $J^y_x$ from linearly polarised along the $x$ (green line) or $y$ (black line) axis light are shown. In (\textbf{g}-\textbf{i}) $J^z_x$ and $J^z_y$ from circularly polarised light (shadowed dashed red and shadowed dotted blue lines, respectively) together with $J^z_y$ from linearly polarised along the $x$ (green line) or $y$ (black line) axis light are shown. In (\textbf{a}, \textbf{b}, \textbf{d}, \textbf{e}, \textbf{g}, \textbf{h}) $\hbar \omega=1.55$\,eV and in (\textbf{b}, \textbf{c}, \textbf{e}, \textbf{f}, \textbf{h}, \textbf{i}) $\Gamma$=25\,meV. In (\textbf{a}, \textbf{c}, \textbf{d}, \textbf{f}, \textbf{g}, \textbf{i}) the calculations are performed at the true Fermi energy level. For all calculations the considered light intensity is $I$=10\,GW/cm$^2$.}
\label{FM_spin_photoc}
\end{figure*}
%\vspace{0.5cm}

The dependence of the spin photocurrents on the position of the Fermi energy  $\mathcal{E}_F$ in the non-magnetic case is examined in Fig.~\ref{NM_spin_photoc}(\textbf{b}) for $\hbar\omega=1.55$\,eV and  $\Gamma$ of 25\,meV. As far as  $J^x_y$ current responses are concerned, their behavior is very similar for  different light polarization cases. In contrast to charge photocurrents, the computed spin currents exhibit  more pronounced peaks, which can be seen for example at $\approx -1.8$\,eV, $\approx +0.2$\,eV and $\approx +1.7$\,eV. By comparing to the non-magnetic band-structure in Fig~\ref{BiAg2_structure}, it is evident that these peaks originate from the regions where the bands become flatter or come closer, and where the number of  electronic states lending themselves to optical transitions is increased. This feature is easier to distinguish in the non-magnetic case as compared to the ferromagnetic one. In general, the response to light linearly polarised along $y$ (black line) is the largest. An exception to this is observed in the regions where two more pronounced peaks appear in the response to light linearly polarised along $x$ (green line). The $J^x_x$ response to circularly polarised light (shadowed dashed red line) behaves differently from the $J^x_y$. We can detect two regions, one below 0\,eV, where the signal has a negative sign, and one above 0\,eV, where the signal has a positive sign. In both cases the signal reaches a steady value of about 200\,$\hbar$/(2e)\,A/m for a wide range of nearly 1\,eV.

The behavior of the spin photocurrents at the true Fermi level for a broadening of $\Gamma=25$\,meV in relation to the laser pulse frequency $\hbar\omega$ is depicted in Fig.~\ref{NM_spin_photoc}(\textbf{c}). We can see that all $J^x_y$ responses exhibit a similar behavior with a very sharp peak at $\approx 0.25$\,eV, while other peaks with smaller magnitude also arise as the laser frequency increases. By comparing with the non-magnetic band-structure in Fig.~\ref{BiAg2_structure}, it is straightforward to attribute these peaks to transitions at the band-crossings which appear near $\pm 0.25$\,eV at the $\Gamma$ and the K points, near $+1$\,eV at the $\Gamma$ point, near $+1.75$\,eV at the $\Gamma$ and the M points, and near $-2$\,eV at the $\Gamma$ point (see also discussion in Sec.~V). The largest in magnitude responses  arise for linearly polarised either along $x$ (green line) or along $y$ (black line) light. As far as the $J^x_x$ response is concerned, it also exhibits peaks which correspond to the same transitions, although the response is around five times smaller in magnitude and opposite in sign.

%{\color{blue} Should we try to comment on the correlation between magnetic and charge currents and non-magnetic spin-currents? For example one can try to comment of the fact that nodal points i.e. points where the signal turns to zero as a function of Gamma, omega or Fermi energy is similar in many cases.}

\subsection{Ferromagnetic spin photocurrents}\label{Ferromagnetic spin photocurrents}

\begin{figure*}[t!]
\begin{center}
\rotatebox{0}{\includegraphics [width=0.9\linewidth]{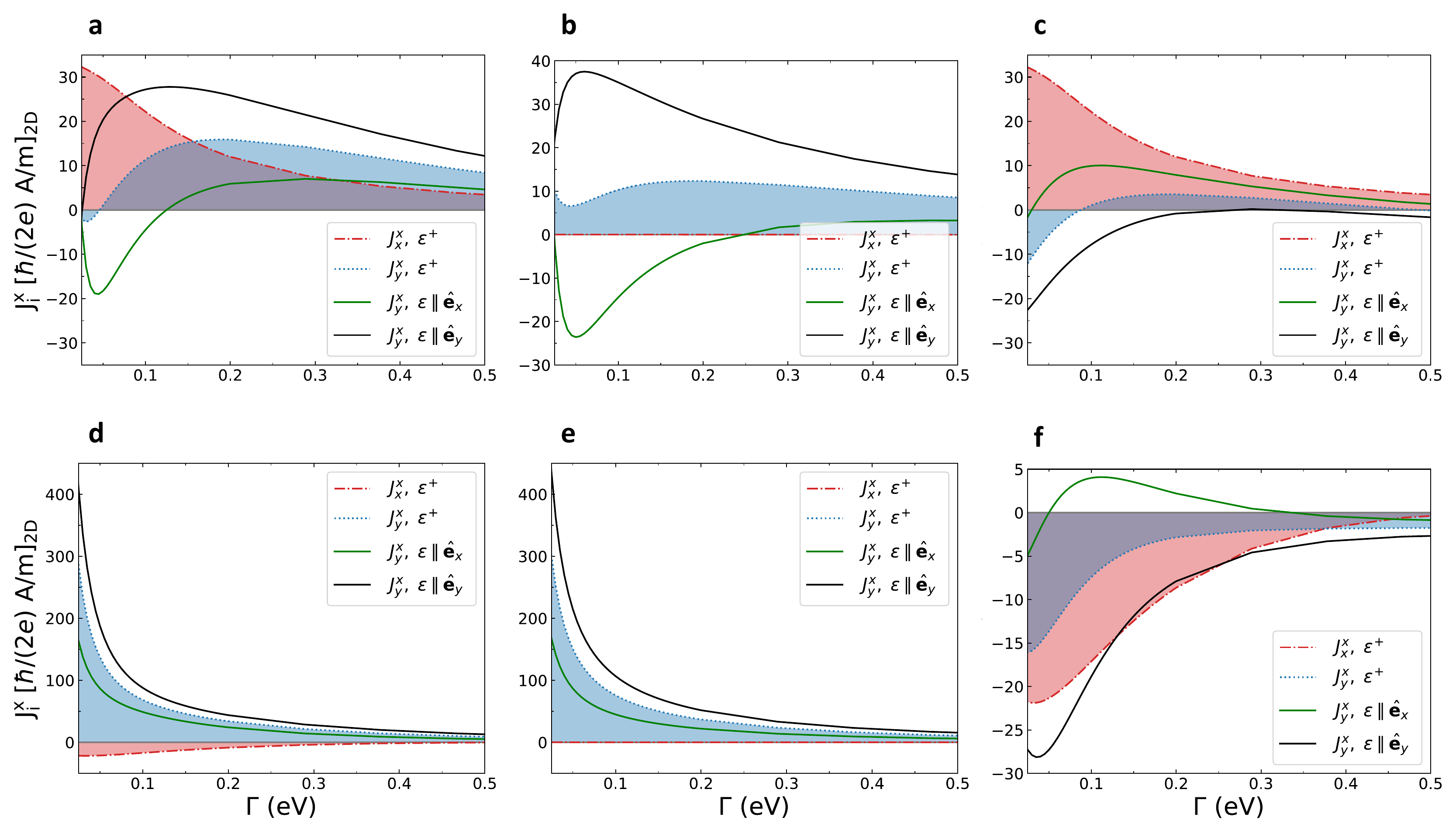}}
\end{center}
\caption{Spin photocurrents in relation to the lifetime broadening $\Gamma$ for the non-magnetic (\textbf{a}-\textbf{c}) and the ferromagnetic (\textbf{d}-\textbf{f}) BiAg$_2$ surface. $J^x_x$ and $J^x_y$ with circularly polarised light (shadowed dashed red and shadowed dotted blue lines, respectively) together with $J^x_y$ linearly polarised along the $x$ (green line) or $y$ (black line) axis are shown. In (\textbf{a}, \textbf{d}) all the contributions to the spin photocurrents are presented, whereas in (\textbf{b}, \textbf{e}) only the two-band and in (\textbf{c}, \textbf{f}) only the three-band contributions are shown. For all cases the calculations are performed at the true Fermi energy level with $\hbar \omega$=1.55\,eV. For all calculations the considered light intensity is $I$=10\,GW/cm$^2$.}
\label{non-res_spin_photoc}
\end{figure*}
%\vspace{0.5cm}

The final case which we examine is that of  laser induced spin-photocurrents for the ferromagnetic BiAg$_2$ surface.  Their dependence concerning the lifetime broadening $\Gamma$ at the true Fermi level and for a laser light frequency of $\hbar\omega=1.55$\,eV is presented in Fig.~\ref{FM_spin_photoc}(\textbf{a}, \textbf{d} and \textbf{g}). For values of $\Gamma$ smaller than 0.25\,eV the behavior of $J^x_i$ and $J^y_i$ becomes non-linear, while $J^z_i$ reaches a peak value and converges to a finite value in the $\Gamma=25$\,meV limit. For the $J^x_y$ response the largest signal arises for linearly polarised along $y$ light [black line in Fig.~\ref{FM_spin_photoc}(\textbf{a})] and for the $J^y_x$ response the largest signal arises for linearly polarised along $x$ light [green line in Fig.~\ref{FM_spin_photoc}(\textbf{d})]. We also point out that $J^x_x$ [shadowed dashed red line in Fig.~\ref{FM_spin_photoc}(\textbf{a})] and $J^y_y$ [shadowed dotted blue line in Fig.~\ref{FM_spin_photoc}(\textbf{d})] current responses to circularly polarised light are rather suppressed in magnitude. Regarding the $J^z_i$ currents, their magnitude is one order of magnitude smaller in comparison to $J^x_i$ and $J^y_i$, with the largest signal appearing for the $J^z_x$ response to circularly polarised light [shadowed dashed red line in Fig.~\ref{FM_spin_photoc}(\textbf{g})].

In Fig.~\ref{FM_spin_photoc}(\textbf{b}, \textbf{e} and \textbf{h}) we present the calculated spin photocurrents in response to variation of the Fermi energy $\mathcal{E}_F$ for  $\hbar\omega=1.55$\,eV and $\Gamma$ of 25\,meV. In general we observe a  larger magnitude of $J^x_y$ current response, with $J^y_x$ response being smaller by a factor of about two. In both cases light linearly polarised along $y$ (black line in Fig.~\ref{FM_spin_photoc}(\textbf{b}) for $J^x_y$ and black line in Fig.~\ref{FM_spin_photoc}(\textbf{e}) for $J^y_x$) is the most optimal choice as it results in larger signal for a wide range of energies. Also, for both cases, narrower regions where the response to light linearly polarised along $x$ (green line in Fig.~\ref{FM_spin_photoc}(\textbf{b}) for $J^x_y$ and green line in Fig.~\ref{FM_spin_photoc}(\textbf{e}) for $J^y_x$) becomes largest can be seen. Interestingly, for currents flowing in the direction of spin polarisation, the response to circularly polarised light (shadowed dashed red line in Fig.~\ref{FM_spin_photoc}(\textbf{b}) for $J^x_x$ and shadowed dotted blue line in Fig.~\ref{FM_spin_photoc}(\textbf{e}) for $J^y_y$) are  suppressed in comparison to  $J^x_y$ and $J^y_x$ currents. As far as the $J^z_i$ currents are concerned, their magnitude is one order of magnitude smaller than in other cases. For all  spin polarization directions we observe that the shape of the curves is very ragged which makes it difficult to attribute given peaks to  certain energy regions of the ferromagnetic band-structure shown in Fig.~\ref{BiAg2_structure}. One can nevertheless notice that $J^y_x$ and $J^z_y$ currents have a similar shape with a change in sign despite their difference in magnitude.

We finally examine spin photocurrents as a function of the  laser frequency, presenting the results  in Fig.~\ref{FM_spin_photoc}(\textbf{c}, \textbf{f} and \textbf{i}). The calculation is once again performed at the true Fermi level $\mathcal{E}_F$ for a lifetime broadening $\Gamma=25$\,meV. For all  spin polarization directions we notice that a very large peak appears near 0.25\,eV and the behavior of the signals continues to be spiky at the same time decreasing in magnitude as the frequency  increases. The peaks near 0.25\,eV and  0.75\,eV can be attributed to the band splittings which appear at the same energies of the ferromagnetic band-structure shown in Fig.~\ref{BiAg2_structure},~i.e.~near $\pm 0.25$\,eV and near $\pm 0.75$\,eV, at the $\Gamma$ and K points, respectively, after the application of the exchange field. For  selected parameters,  $J^x_y$ and $J^y_x$ currents appear to be similar in magnitude while $J^z_i$ currents are one order of magnitude smaller. Also, once again the $J^x_x$ [shadowed dashed red line in Fig.~\ref{FM_spin_photoc}(\textbf{c})] and $J^y_y$ [shadowed dotted blue line in Fig.~\ref{FM_spin_photoc}(\textbf{f})] responses to circular light are suppressed as compared to the others. In Fig.~\ref{FM_spin_photoc}(\textbf{c}) the response of $J^x_y$ to linearly polarised light along $x$ (black line) is the largest, while in Fig.~\ref{FM_spin_photoc}(\textbf{f}) the largest response of $J^y_x$ occurs for linearly polarised light either along $x$ (green line) or $y$ (black line).

\section{Two-band vs three-band transitions}\label{two-band_vs_three-band}

In the last section we perform an analysis of decomposition of the photocurrents into two-band and three-band contributions.
In our notation, which follows the Keldysh non-equilibrium formalism, Eq.~\eqref{eq:conductivity} involves a summation over three band indices $-$ $n, m, m'$ $-$ originated from the expansion of three Green functions in the basis of Bloch states,  Eq.~\eqref{eq:green}. A resulting form of the corresponding expression can be seen in Eqs.\,(B4-B6) of Appendix B in Ref.~\cite{freimuth_2016}. In this context, the two-band transitions correspond to the case when $n=m$ while the three-band transitions correspond to the case when $n\neq m$. This is 
 similar to the case considered in Ref.~\cite{Zhang_2018, Zhang_2019}, where computed charge photocurrents are decomposed into two-band and three-band transitions between the energy bands within the second order Kubo formalism~\cite{KrautBaltz_1981}. The former kind of transitions can be considered as resonant transitions between two states that differ from each other by $\pm\hbar\omega$, and the latter as virtual transitions aided by a third band.%, where $n, m$ label the energy bands in Eqs. (B4-B6) in Appendix B of~\cite{freimuth_2016}.

\begin{figure*}[t!]
\begin{center}
\rotatebox{0}{\includegraphics [width=0.9\linewidth]{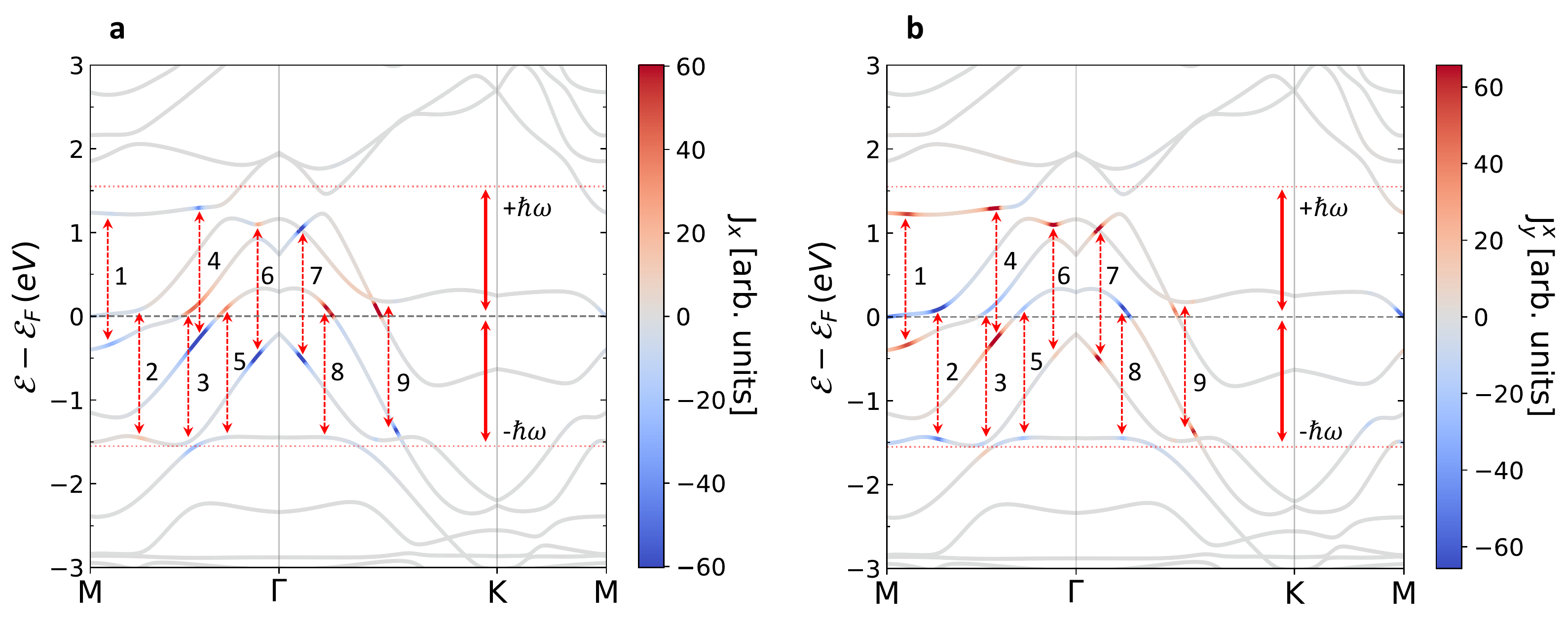}}
\end{center}
\caption{Band-resolved two-band contributions of the laser-induced charge (\textbf{a}) and spin (\textbf{b}) photocurrents for the ferromagnetic BiAg$_2$ surface. Presented in (\textbf{a}) is the $J_x$ response and in (\textbf{b}) is the $J^x_y$ response, both for light linearly polarised along the $x$ axis. The horizontal dotted red lines at $\pm 1.55$\,eV denote the energy of the laser pulse. The dashed red arrows depict two-band transitions between bands whose energy difference matches the laser pulse energy. In both cases $\hbar\omega=1.55$\,eV and $\Gamma=25$\,meV. Throughout the calculations the considered light intensity is $I=10$\,GW/cm$^2$.}
\label{res_charge_spin_photoc_bandplots}
\end{figure*}

In the inset of Fig.~\ref{FM_charge_photoc}(\textbf{a}) the contribution from the three-band transitions to  the laser-induced charge photocurrents in the ferromagnetic BiAg$_2$ surface is shown. We find that for $J_x$ currents the two-band transitions dominate by one to two orders of magnitude, regardless of the light polarisation. This finding is important, since
taking only two-band transitions into account reduces the computational effort significantly.
Moreover, while we also find that two-band $J_x$ currents are odd in the lifetime broadening $\Gamma$ (i.e.~odd in the relaxation time $\tau$ within the constant relaxation time approximation), the $J_y$ response to circularly polarised light consists only of three-band transitions which are even in the lifetime broadening.
Furthermore, in Fig.~\ref{non-res_spin_photoc}(\textbf{a}-\textbf{c}) we present the spin photocurrents for the non-magnetic case of the BiAg$_2$ surface, decomposed into two-band and three-band contributions. For the $J^x_y$ current response, it is evident that  both types of transitions contribute with similar magnitude to the resulting signal, and we also find that  these currents are odd in the lifetime broadening $\Gamma$. On the contrary,  $J^x_x$ currents arising in response to circularly polarised light are driven exclusively by three-band transitions, exhibiting at the same time an even $\Gamma$ dependence.

For the ferromagnetic spin photocurrents on the BiAg$_2$ surface we present in Fig.~\ref{non-res_spin_photoc}(\textbf{d}-\textbf{f}) total, two-band and  three-band contributions, respectively, to $J^x_i$ currents. Similarly to the ferromagnetic $J_x$ charge photocurrents, we notice that the two-band contribution is the most significant one for the $J^x_y$ current responses. Comparing to the ferromagnetic $J_x$ charge photocurrents, we find that three-band transitions are larger in this case but still not important for the whole effect. In addition, we predict that $J^x_y$ currents are odd in $\Gamma$. On the other hand, the response of $J^x_x$ to circularly polarised light consists only of three-band transitions and it is even in $\Gamma$. %Furthermore, for simplicity we choose not to show this analysis for the $J^y_i$ and $J^z_i$ ferromagnetic spin photocurrents and simply discuss them in the following. 
The case for $J^y_i$ (not shown) is similar to $J^x_i$ with  $J^y_x$ response mediated by two-band transitions and $J^y_y$ current response to circularly polarised light mediated by three-band transitions, while being odd and even in $\Gamma$, respectively. Finally, all $J^z_i$ currents comprise only three-band transitions, with the $J^z_x$ being odd and  $J^z_y$ being even in $\Gamma$. The important conclusion that we can draw from our calculations is that in case of spin currents it is generally difficult to attribute a given component of the spin photocurrent to a specific type of band transitions without an in-depth analysis of the symmetry or explicit calculations.

In a last step, we calculate the band-resolved laser-induced charge and spin photocurrents for the ferromagnetic BiAg$_2$ surface by taking into account only the dominant contributions from the two-band transitions. The former is presented in Fig.~\ref{res_charge_spin_photoc_bandplots}(\textbf{a}) and the latter in Fig.~\ref{res_charge_spin_photoc_bandplots}(\textbf{b}). For both cases we notice the appearance of ``hotspots" where the magnitude of the photocurrents reaches large values as compared to the rest of the considered region where they vanish, while they are located within the energy region which is affected by the laser excitation, i.e $[\mathcal{E}_f-\hbar\omega, \mathcal{E}_f+\hbar\omega]$. Moreover, we are able to classify the ``hotspots" in pairs which are located at the same position in $k$-space and their energy difference is equal to the energy of the laser pulse. These can be interpreted as two-band transitions which are generated by the laser-excitation, as highlighted with numbered dashed red arrows in Fig.~\ref{res_charge_spin_photoc_bandplots}. We notice that these transitions are common both for charge and spin photocurrents but only a few of them appear equally pronounced in both cases (see arrows 4, 7 and 9 in Fig.~\ref{res_charge_spin_photoc_bandplots}). On the other hand there are transitions in which either the upper or the lower band appears pronounced (see arrows 3, 5, 6 and 8 in Fig.~\ref{res_charge_spin_photoc_bandplots}). There is also the case of transitions which are very prominent for the spin currents but for the charge currents their magnitude is small (see arrows 1 and 2 in Fig.~\ref{res_charge_spin_photoc_bandplots}).
% We notice that most of these transitions are common both for charge and spin photocurrents (see arrows 1 $-$ 5 in Fig.~\ref{res_charge_spin_photoc_bandplots}) and some are active only for the spin photocurrents (see arrows 6 $-$ 9 in Fig.~\ref{res_charge_spin_photoc_bandplots}). %, i.e. the two transitions close the M point.
For the charge photocurrents the observed transitions either maintain the sign of the photocurrents [see arrows 2 and 4 in Fig.~\ref{res_charge_spin_photoc_bandplots}(\textbf{a})] or change it [see arrows 1, 3 and 5 in Fig.~\ref{res_charge_spin_photoc_bandplots}(\textbf{a})]. In addition, no change of sign is observed for the spin photocurrents. In general, all the states participating in two-band transitions have predominantly Bi $p$-character, thus making these states the only ones which contribute to the response effects in our system. This fact highlights even more the importance of the joint effect of Rashba and exchange splitting on the generation of the charge and spin photocurrents. 
%Similar band-resolved plots for the non-magnetic case or for the three-band contributions are not presented because they don't carry important information about the generation of the photocurrents.
%{\color{red} I didn't add the spin-polarised bands. The only thing that I noticed is that in Fig.6a the transitions were the sign changes (1, 3, 5) are between bands with different spin character.}

%\cleardoublepage

\section{Summary}\label{Summary}

By using Keldysh formalism, we performed first-principles calculations of charge and spin photocurrents which arise as a second-order response to laser excitation at a BiAg$_2$ surface.
%Our results, both for the non-magnetic and the ferromagnetic cases, are in full agreement with the predictions of the Rashba model concerning the appearance of these photocurrents and their symmetry properties regarding the light helicity and the magnetization direction. 
We find that the calculated responses are large in magnitude and depend strongly on such material parameters as the lifetime broadening or the fine details of the electronic structure such as the position of the Fermi energy level and the laser pulse energy. These findings mark the surfaces and interfaces of strongly spin-orbit coupled materials as efficient sources not only of planar charge photocurrents but also currents of spin, whose role in the processes of THz emission by spintronics systems has been possibly overlooked.
%In general the largest signals appear as a response to linearly polarised light along $x$ or $y$. 
%We find also also that the computed charge and spin photocurrents are similar in magnitude and that the pronounced variations which appear are clear fingerprints of the electronic structure. 
Overall, our calculations for the experimentally well-studied BiAg$_2$ surface can be extremely useful for understanding the role of the interfacial Rashba spin-orbit interaction in generating various photocurrents. As such our calculations can also serve as a benchmark for further material-specific studies of photo-induced charge and spin dynamics.

%\noindent
%{\it Acknowledgements.}
\section{Acknowledgements}
%We thank Julen Azpiroz for extensive discussions on the subject. 
This work was supported by the Deutsche Forschungsgemeinschaft (DFG, German Research Foundation) $-$ TRR 173/2 $-$ 268565370 (project A11), TRR 288 – 422213477 (project B06), and the Sino-German research project DISTOMAT (MO 1731/10-1). This project has received funding from the European Union’s Horizon 2020 research and innovation programme under the Marie Skłodowska-Curie grant agreement No 861300.
%, and under synergy grant "3D MAGiC” agreement (Grant No. 856538).
%We  acknowledge  funding  under SPP 2137 ``Skyrmionics" of the DFG.
%(project  MO  1731/7-1) 
%and project MO 1731/5-1 
% We gratefully acknowledge financial support from the European Research Council (ERC) under the European Union's Horizon 2020 research and innovation program (Grant No. 856538, project "3D MAGiC”).
We  also gratefully acknowledge the J\"ulich Supercomputing Centre and RWTH Aachen University for providing computational resources under projects  jiff40 and jara0062.

%----------------------------------------------------
% literature
%----------------------------------------------------

\hbadness=99999 
\bibliographystyle{apsrev4-2}%{naturemag}%
\bibliography{BiAg2_photoc}

% %----------------------------------------------------
% \cleardoublepage
% \appendix
% %----------------------------------------------------

% \onecolumngrid

% \begin{center}
%     {\bf \large Supplemental Material: The chiral Hall effect of magnetic skyrmions from a general cyclic cohomology approach}
% \end{center}

\end{document}